# Trust of Strangers: a framework for analysis

Shawn Berry, DBA[1]*

**January 4, 2025**

[1]William Howard Taft University, Lakewood CO, USA

*Correspondence: shawnpberry@gmail.com

**Abstract** Trust among people is essential to ensure collaboration, social network building, transactions, and the development and engagement of new audiences for brand promotion or social causes. In Berry (2024), the trust attitudes of respondents toward strangers on the street, other groups of people, and information sources were measured. This study evaluates the trust of strangers using a 5-factor structural equation model. The analysis yielded a robust model with four of five factors and all variables being statistically significant, with social trust and institutional trust yielding the greatest positive effect on trust of strangers on the street. While demographic characteristics had a small positive effect, the trust of friends and family had a mild negative effect on the trust of strangers on the street. Trust of information sources was not statistically significant and had a negligible positive effect on the trust of strangers. The results also indicate that almost 48% of respondents distrust strangers on the street, implying that trust is not automatically endowed. Directions for future research and implications for business and social causes are discussed.

**Keywords:** trust, distrust, strangers, social trust, generalized trust model

## 1. Introduction

The concept of trust has been studied in various contexts, such as the willingness to help strangers or those in need (De Jong, Van der Vegt, & Molleman, 2007; Seo et al., 2012; ), the trust of supervisors and helping colleagues at work (Poon, 2006), teammates and teamwork (Spector & Jones, 2004), how trust affects workplace performance (Snow, 1996; Addison & Teixeira, 2020), and the trust of coworkers and organizations (Tan and Lim, 2010; Bachmann, Gillespie & Priem, 2015). The measurement of trust using questionnaires has its origins with Cantril and Strunk (1951). Since then, models of generalized trust have evolved which take into account multiple factors that shape the trust given to people (Rothstein & Stolle, 2001; Freitag & Bauer, 2016; Robbins, 2022). According to Pew Research Center (2019), among Americans, 52% of respondents said that "most can be trusted" and 47% of respondents said that "most can't be trusted". Research carried out by Edelman on the topic of trust includes the measurement of attitudes towards institutions, government, and information sources. In its annual Edelman Trust Barometer, the company reports trends of trust in these categories, and their implications for society to carry on important work and solve problems, such as vaccine hesitancy, for example. Insights from the 2024 Edelman Trust Barometer reveals that sources of credible information that people trust are, in order from most credible to least credible were "a person like yourself, academic expert, company technical expert, NGO representative, journalist, regular employee, CEO, board of directors, government official" (Edelman, 2024, p.25). In fact, the majority of Americans feel that the media is biased, suggesting that there is deliberate misinformation, that the media is not objective, and news outlets to be partisan in nature (Edelman, 2021, p.26). This suggests that information sources are incredibly important to the formation of trust. Edelman offers a general portrait of the U.S. population as it relates to high and low trust, stating:

"high trusters are 22 percent of the population, low trusters are 35 percent of the population and medium trusters are 41 percent of the population. The most trusting groups are people with postgraduate education, whites, $75,000-plus income and over 65 years in age. The least trusting groups are people with high school or less, blacks and Hispanics, under $30,000 income, and 18-to-29 year-olds" (Edelman, 2019).

Edelman (2021) reported that, in order of trust from highest to lowest, Americans trusted business, nongovernmental organizations, media, and government. (p.6). All of these insights serve to inform that trust development among people is also influenced by our trust of organizations and sources of information.

The factors that affect trust are varied, and the interrelationships among variables and factors must be closely examined. For this reason, scholars have identified that approaches to modeling trust are required (Cho, Chan & Adali, 2015). To this end, attempts have been made to measure trust. One such model by Naef and Schupp (2009) proposed a fount-point Likert measurement scale for the trust of strangers and they "show that the dimension of this scale is distinct from trust in institutions and trust in known others" (p.1). However, the authors state that "the experimental measure of trust is, on the other hand, not significantly correlated with trust in institutions nor with trust in known others" (p.1). Frazier, Johnson, and Fainshmidt (2013) proposed a propensity to trust scale that leveraged constructs of trustworthiness, trust, and optimism, and evaluated aspects of the integrity, ability, and benevolence of respondents. While their particular model was validated across four studies, the authors note that "there has been comparatively little research on propensity to trust in the literature" (Frazier, Johnson, and Fainshmidt, 2013).

The trust of strangers on the street will be explored in this study. Using data from my dissertation (Berry, 2024), trust of strangers will be analyzed according to the demographic characteristics of respondents with respect to age, gender, level of education, level of income, and region of the United States to get a broad picture. More specifically, these trust levels will be analyzed with structural equation modeling (SEM) using constructs of trust of information, media, business and persons. Using these insights, new knowledge with respect to the trust of strangers can be added which are of value to marketing and sociology researchers. The study will conclude with recommendations and directions for future research.

## 2. Materials and Methods

The data for this study was collected using Amazon Mechanical Turk (mTurk), administered by way of an online questionnaire that was used for my dissertation (Berry, 2024). The original sample consisted of 398 male and female mTurk users living in the United States. The respondent data collected consisted of information concerning habits, attitudes, beliefs, and non-identifying demographic characteristics that relate to their online review posting and reading activities (Berry, 2024). Of the original sample size, "11.8% of the 398 respondents were excluded due to not consenting to the study (3), self-reporting as not residing in the United States (8), attention check question failure (24), and identification of attempts to take the survey more than once (12)." (p.3), and the final sample was n=351 (Berry, 2024).

Trust levels among respondents were evaluated using responses to two instruments that quantify the trust of people and the trust of information sources in Berry (2024) by rating a collection of individuals in varying roles in society with a 5-point Likert scale (Berry, 2024). The trust of people instrument consisted of 13 items where respondents evaluated a variety of different individuals in society (e.g., family, friends, salespeople, new immigrants, strangers, etc.). The trust of people instrument was evaluated with Cronbach's alpha, the value of which "was 0.85 (Berry, 2024), which is considered to be high (Taber, 2018), and therefore, reliable" (Berry, 2024, p.3). The trust of information instrument consisted of 10 items where respondents evaluated their level of trust of different online and paper-based information sources (e.g., periodicals, online discussion forums, etc.). The trust of information instrument was also evaluated with Cronbach's alpha, the value of which "was 0.92 and can be classified as high or reliable (Taber, 2018)" (Berry, 2024, p.50).

The data was coded for analysis according to the scheme as illustrated in Table 1 below.

**Table** 1

*Variable coding scheme*

| Variable name | Variable description | Variable type | Coding |
|---|---|---|---|
| Age | Age group of respondent, years of age | Categorical | 18-24 = 1<br>25-34 = 2<br>35-44 = 3<br>45-54 = 4<br>55 and over = 5 |
| Gender | Gender of respondent | Categorical | Female = 0<br>Male = 1<br>Non-binary = 2 |
| Income | Annual income level of respondent, USD | Categorical | Less than $30,000 = 1<br>$30,000-$49,999 = 2<br>$50,000-$69,999 = 3<br>$70,000 and over = 4 |
| Education | Education level of respondent | Categorical | Did not finish high school = 0<br>High school graduate = 1<br>Some college = 2<br>Bachelor's degree = 3<br>Master's degree = 4<br>Post-graduate or higher = 5 |
| Region | United States region of residence of respondent | Categorical | Middle Atlantic = 1<br>New England = 2<br>South Atlantic = 3<br>East South Central = 4<br>West South Central = 5<br>Mountain = 6<br>Pacific = 7 |
| Likert scores | Degrees of agreement or importance of behavioral factors to measure trust | Ordinal | Definitely distrust = 1<br>Somewhat distrust = 2<br>Neither trust nor distrust = 3<br>Somewhat trust = 4<br>Definitely trust = 5 |

| | of people | | |
|---|---|---|---|
| Everlied | If respondent admits to ever have posted a fake online review | Binary | Has not = 0<br>Has = 1 |

**Source:** Berry (2024), Table 1.

Using the trust level data collected through the instruments and constructs used in Berry (2024), a five-factor conceptual framework for the trust of strangers on the street was created. The conceptual framework is illustrated in Figure 1. Each factor in the framework acts as a source of influence that affects the trust of strangers on the street. In this framework, the extent to which respondents trust various kinds of people, and sources of influence, such as various information sources and personalities or celebrities in society, ultimately influences the extent to which strangers are trusted. Personal trust consists of trust attitudes toward people who are particularly close in everyday life to the respondent. Social trust consists of trust attitudes toward people that respondents probably do not personally know but may be influenced by, either as fans or followers, or as spokespeople for a product or cause. Institutional trust consists of trust attitudes toward representatives of government, religion, and business owners as a proxy for trust of organizations as sources of personal influence. Information trust consists of trust attitudes toward various sources of information, including forums where advice and feedback is given by other anonymous and random strangers and relied upon for decisions. Finally, demographics are used as a factor to determine whether certain characteristics of respondents activate trust of strangers, with the variable everlied as a proxy for honesty.

**Figure 1**

*Conceptual framework*

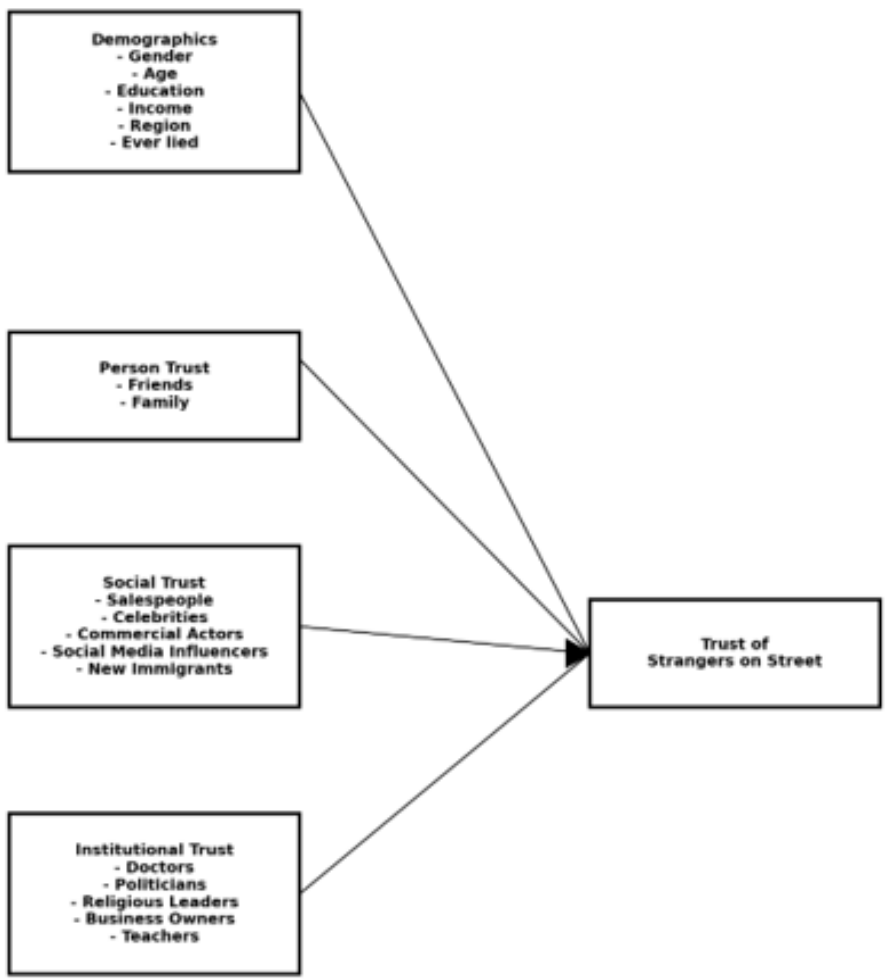

The items for each trust instrument from Berry (2024) were allocated to new constructs in this current framework to study the trust of strangers on the street. Table 2 below illustrates how each item of their respective constructs from Berry (2024) were allocated to each of the constructs in the framework. The demographic indicators were collected at the end of the questionnaire in Berry (2024), and therefore, are individual questions designed to collect information about the respondents age, gender, region of residence, education level, and income level.

**Table 2**

Framework constructs, variables, and trust instrument data source

| Framework construct | Indicator | Trust instrument data source (Berry, 2024) |
|---|---|---|
| Person Trust | family | Trust of people |
| Person Trust | friends | Trust of people |
| Institutional Trust | doctors | Trust of people |
| Institutional Trust | politicians | Trust of people |
| Institutional Trust | religious | Trust of people |
| Institutional Trust | businessowners | Trust of people |
| Institutional Trust | teachers | Trust of people |
| Social Trust | salespeople | Trust of people |
| Social Trust | celebrities | Trust of people |
| Social Trust | commercialactors | Trust of people |
| Social Trust | socmediainfluencers | Trust of people |
| Social Trust | newimmigrants | Trust of people |
| Information Trust | Buswebsite | Trust of information |
| Information Trust | BusFacebookpage | Trust of information |
| Information Trust | periodicals | Trust of information |
| Information Trust | internetsearch | Trust of information |
| Information Trust | googlebuslisting | Trust of information |
| Information Trust | internetdiscussforum | Trust of information |
| Information Trust | consumerratingssite | Trust of information |
| Demographics | age | Respondent demographics |
| Demographics | gender | Respondent demographics |
| Demographics | income | Respondent demographics |
| Demographics | education | Respondent demographics |
| Demographics | region | Respondent demographics |
| Demographics | everlied | Question: Have you ever posted an online review that was not true? |

**Source:** Berry (2024)

The literature surrounding the trust of strangers on the street suggests several consistent relationships. Certain demographic characteristics have been recognized as having a positive effect on the trust of strangers. Almakaeva, Welzel, & Ponarin (2018) observed that "in countries with advanced human empowerment, a much broader set of individual-level characteristics increases trust in strangers. This set includes ethnic tolerance, membership in voluntary associations, social movement activity, emancipative values, subjective well-being, age, and education" (p.923). In his study of trust in Iran, Mirfardi (2011) concluded "that there is significant relationship between all of independent variables (Gender, age, education level, job situation, marital situation) and social trust to families and relatives, there is significant relationship between variables such as gender, education level, job situation, marital situation (independent variables) and social trust to friends" (p.168). Alesina and La Ferrara (2002) found that "find that the strongest factors associated with low trust are… (iii) being economically unsuccessful in terms of income and education" (p.207), adding that "religious beliefs and ethnic origins do not significantly affect trust" (p.207). Ermisch et al. (2009), in their study of the trust of strangers in Britain, "suggests that trusting is more likely if people are older, their financial situation is either 'comfortable' or 'difficult' compared with 'doing alright' or 'just getting by'" (p.749). The role of institutional trust as a factor for trusting strangers was examined by Kaina (2011) who posited that the trust of strangers is based on the belief that others will think and act similarly, given "the potential of institutions for enabling people to trust strangers rests on institutions' power to structure individual action. The endurance and efficacy of institutions rather than their normative principles give us significant clues that our anonymous fellow citizens think about institutions as we do and accordingly feel committed to the rules of action" (p.282).

## 3. Results

The descriptive statistics are shown in Table 3. Table 3 suggests that the mean trust level of respondents for strangers on the street is only marginally higher than the mean trust level of celebrities and social media influencers and just below the mean trust of new immigrants. The median trust level of strangers was 3 or neither trust nor distrust, suggesting indifference or perhaps implying ambivalence toward strangers (Berry, 2024). Ironically, as interpersonal sources of influence and trust, respondents trusted the opinions of strangers posting in internet discussion forums and consumer ratings sites more than a stranger on the street. While salespeople and politicians were the least trusted kinds of people, family and friends were the most trusted kinds of people. Thus, respondents place greater trust in the feedback of others and certain figures in society that they depend upon or directly interact with, such as teachers and doctors. Less trust is endowed to those who appear to spokespersons or representatives who may have a particular goal or agenda, such as social media influencers and politicians, and not as personally close as family, friends, and those that know an individual. Therefore, the trust of strangers serves as a comparison benchmark for the extent to which individuals are willing to trust others, and subsequently allow them to influence individuals.

**Table 3**

Descriptive statistics

| Variable | Mean | SD | Median | SE | Min | Max | Range | Skewness | Kurtosis |
|---|---|---|---|---|---|---|---|---|---|
| Everlied | 0.17 | 0.37 | 0 | 0.02 | 0 | 1 | 1 | 1.79 | 1.23 |
| Business website | 3.3 | 1.14 | 3 | 0.06 | 1 | 5 | 4 | -0.3 | -0.63 |
| Business Facebook page | 3.19 | 1.03 | 3 | 0.05 | 1 | 5 | 4 | -0.29 | -0.47 |
| Periodicals | 3.08 | 0.99 | 3 | 0.05 | 1 | 5 | 4 | -0.02 | -0.17 |
| Internet search | 3.39 | 0.97 | 4 | 0.05 | 1 | 5 | 4 | -0.46 | -0.17 |
| Google business listing | 3.7 | 0.99 | 4 | 0.05 | 1 | 5 | 4 | -0.9 | 0.54 |
| Internet discussion forum | 3.38 | 1.09 | 4 | 0.06 | 1 | 5 | 4 | -0.39 | -0.49 |
| Consumer ratings site | 3.57 | 1.01 | 4 | 0.05 | 1 | 5 | 4 | -0.67 | -0.07 |
| Friends | 4.2 | 0.97 | 4 | 0.05 | 1 | 5 | 4 | -1.38 | 1.67 |
| Family | 4.27 | 0.94 | 4 | 0.05 | 1 | 5 | 4 | -1.57 | 2.46 |
| Doctors | 3.7 | 1.04 | 4 | 0.06 | 1 | 5 | 4 | -0.73 | 0.13 |
| Politicians | 2.16 | 1.1 | 2 | 0.06 | 1 | 5 | 4 | 0.63 | -0.5 |
| Salespeople | 2.34 | 1.09 | 2 | 0.06 | 1 | 5 | 4 | 0.54 | -0.37 |
| Strangers on street | 2.53 | 1.05 | 3 | 0.06 | 1 | 5 | 4 | 0.2 | -0.54 |
| New immigrants | 2.76 | 1.01 | 3 | 0.05 | 1 | 5 | 4 | 0.11 | 0.01 |
| Business owners | 3.02 | 0.88 | 3 | 0.05 | 1 | 5 | 4 | 0.07 | -0.06 |
| Celebrities | 2.49 | 1.08 | 2 | 0.06 | 1 | 5 | 4 | 0.51 | -0.29 |
| Commercial actors | 2.32 | 1.15 | 2 | 0.06 | 1 | 5 | 4 | 0.59 | -0.39 |
| Religious | 3.02 | 1.2 | 3 | 0.06 | 1 | 5 | 4 | -0.07 | -0.78 |
| Social media influencers | 2.43 | 1.11 | 2 | 0.06 | 1 | 5 | 4 | 0.3 | -0.84 |
| Teachers | 3.51 | 1 | 4 | 0.05 | 1 | 5 | 4 | -0.49 | -0.14 |
| Gender | 0.32 | 0.49 | 0 | 0.03 | 0 | 2 | 2 | 0.97 | -0.53 |
| Age | 2.7 | 0.96 | 3 | 0.05 | 1 | 5 | 4 | 0.38 | -0.36 |
| Education | 2.58 | 1.03 | 2 | 0.06 | 0 | 5 | 5 | 0.29 | -0.29 |
| Income | 2.18 | 1.13 | 2 | 0.06 | 1 | 4 | 3 | 0.48 | -1.16 |
| Region | 3.75 | 1.92 | 4 | 0.1 | 1 | 7 | 6 | 0.15 | -0.96 |

**Source: data analysis**

Table 4 illustrates the overall distribution of the trust levels toward strangers on the street among respondents according to the Likert scale scores. While the distribution suggests that just over a third of respondents neither trust nor distrust strangers, nearly half of the respondents express distrust of strangers. Only 16.24% of respondents expressed trust of strangers, with just over 3% expressing that they definitely trust strangers. This distribution generally implies that respondents do not explicitly trust strangers. Moreover, if those respondents that declare to neither trust nor distrust strangers are assumed to effectively mean that they do not bestow trust, the distribution of trust shown in the table implies that almost 84% of people do not trust strangers on the street.

**Table 4**

*Distribution of trust of strangers on the street by respondents*

| Trust Level | Frequency | Percentage | Cumulative % |
|---|---|---|---|
| 1 - Definitely distrust | 67 | 19.09 | 19.09 |
| 2 - Somewhat distrust | 100 | 28.49 | 47.58 |
| 3 - Neither trust nor distrust | 127 | 36.18 | 83.76 |
| 4 - Somewhat trust | 45 | 12.82 | 96.58 |
| 5 - Definitely trust | 12 | 3.42 | 100.00 |
| Total | 351 | 100 | 100 |

**Source: data analysis**

The trust of strangers on the street among respondents was analyzed with respect to the age category. The data are presented in Table 5. The age category with the greatest distrust of strangers is among those respondents aged 25 to 34 years (19.7% of grand total, 41.3% of all respondents indicating definitely and somewhat distrust of strangers). Those in younger age categories appear to distrust strangers more than those in older age categories (45 and older). In general, distrust of strangers is greater than trust of strangers across all age categories. 36.2% of all respondents neither distrust nor trust strangers. Only 16.2% of respondents trust strangers, whether somewhat and definitely trust. Chi-square analysis was performed on the data. There is no statistically significant relationship between the trust of strangers on the street and age category $X^2$(16, N = 351) = 10.06, p = .863.

**Table 5**

*Respondent age distribution of trust of strangers on street*

| Age | 1 Definitely distrust | 2 Somewhat distrust | 3 Neither trust nor distrust | 4 Somewhat trust | 5 Definitely trust | Totals |
|---|---|---|---|---|---|---|
| 18–24 | 5 | 6 | 13 | 1 | 1 | 26 |
| 25–34 | 26 | 43 | 49 | 16 | 5 | 139 |
| 35–44 | 23 | 34 | 38 | 15 | 4 | 114 |
| 45–54 | 12 | 12 | 23 | 11 | 1 | 59 |
| 55 and older | 1 | 5 | 4 | 2 | 1 | 13 |
| **Totals** | 67 | 100 | 127 | 45 | 12 | 351 |

**Source: data analysis**

The trust of strangers on the street among respondents was analyzed with respect to self-reported untrue online review posting behavior as a proxy for the honesty of a respondent. The data are presented in Table 6. The data suggests that those that have admitted to posting untrue online reviews appear to have higher trust levels toward strangers on the street. While, the majority of people (83.5%) admit to never having posted an untrue online review, 52.9% of this group definitely distrust and somewhat distrust strangers on the street. The Chi-square analysis was performed on the data. There is a statistically significant relationship between the trust of strangers on the street and untrue online review posting behavior by respondents $X^2(4, N = 351) = 88.155, p < .001$.

**Table 6**

*Respondent distribution of trust of strangers on the street versus untrue online reviews posting behavior*

| Have you ever posted an online review that was untrue | 1 - Definitely distrust | 2 - Somewhat distrust | 3 - Neither trust nor distrust | 4 - Somewhat trust | 5 - Definitely trust | Totals |
|---|---|---|---|---|---|---|
| Never posted an untrue online review | 62 | 93 | 114 | 21 | 3 | 293 |
| Has posted an untrue online review | 5 | 7 | 13 | 24 | 9 | 58 |
| **Totals** | 67 | 100 | 127 | 45 | 12 | 351 |

**Source: data analysis**

Table 7 below illustrates the distribution of trust of strangers on the street according to gender. with respect to the level of education of the respondents. Males tended to distrust strangers on the street about the same as females (47.2%). 19.4% of females tended to somewhat trust and definitely trust strangers as opposed to 15% of males. 37.1% of males neither trust nor distrust strangers on the street as compared with 33.3% of females. The Chi-square analysis was performed on the data. There is no statistically significant relationship between the trust of strangers on the street and gender $X^2(8, N = 351) = 4.763, p = .783$.

**Table 7**

*Respondent distribution of trust of strangers on the street according to gender*

|  | 1 - Definitely distrust | 2 - Somewhat distrust | 3 - Neither trust nor distrust | 4 - Somewhat trust | 5 - Definitely trust | Totals |
|---|---|---|---|---|---|---|
| Male | 44 | 71 | 89 | 27 | 9 | 240 |
| Female | 23 | 28 | 36 | 18 | 3 | 108 |
| Non-binary | 0 | 1 | 2 | 0 | 0 | 3 |
| **Totals** | 67 | 100 | 127 | 45 | 12 | 351 |

**Source: data analysis**

Table 8 below illustrates the distribution of trust of strangers on the street with respect to the level of education of the respondents. The greatest distrust of strangers on the street was observed among those respondents possessing some college education and with bachelor's degrees (35.3% of grand total, 74.3% of respondents that definitely and somewhat distrust strangers). Among respondents that have a master's degree or higher appear to trust and distrust strangers equally. Those respondents that did not have any college education or degree appear to generally distrust strangers more than trust strangers. Indifference toward strangers was also observed among those respondents with some college education and with bachelor's degrees (24.8% of grand total, 68.5% of those respondents that neither trust nor distrust). Chi-square analysis was performed on the data. There is a statistically significant relationship between the trust of strangers on the street and education level $X^2(20, N = 351) = 47.09, p = .0006$.

**Table 8**

*Respondent education level distribution of trust of strangers on the street*

| Level of education | 1 Definitely distrust | 2 Somewhat distrust | 3 Neither trust nor distrust | 4 Somewhat trust | 5 Definitely trust | Totals |
|---|---|---|---|---|---|---|
| Did not finish high school | 2 | 0 | 0 | 0 | 0 | 2 |
| High school graduate | 7 | 11 | 21 | 6 | 1 | 46 |
| Some college | 26 | 40 | 51 | 8 | 3 | 128 |
| Bachelor's degree | 23 | 35 | 36 | 15 | 3 | 112 |
| Master's degree | 9 | 10 | 10 | 15 | 5 | 49 |
| Postgraduate or higher | 0 | 4 | 9 | 1 | 0 | 14 |
| **Totals** | 67 | 100 | 127 | 45 | 12 | 351 |

**Source: data analysis**

Table 9 illustrates the distribution of the trust of strangers on the street according to the income level of the respondents. The greatest distrust of strangers on the street was observed among respondents that earned $49.999 per year or less (31.6% of grand total, 66.5% of respondents that definitely and somewhat distrust strangers). By comparison, 77.2% of respondents that somewhat and definitely trust strangers earned $49,999 per year or less (12.5% of grand total). The greatest trust of strangers was observed among respondents earning between $30,00 and $49,999 per year (8% of grand total, 45.6% of respondents that definitely and somewhat trust strangers). In general, across all income categories, more respondents expressed distrust than trust toward strangers on the street. Chi-square analysis was performed on the data. There is a statistically significant relationship between the trust of strangers on the street and income level $X^2(12, N = 351) = 24.41, p = .018$.

**Table 9**

*Respondent income level distribution of trust of strangers on the street*

| Income level | 1 Definitely distrust | 2 Somewhat distrust | 3 Neither trust nor distrust | 4 Somewhat trust | 5 Definitely trust | Totals |
|---|---|---|---|---|---|---|
| Less than $30,000 | 19 | 46 | 45 | 11 | 5 | 126 |
| $30,000–$49,999 | 20 | 26 | 32 | 23 | 5 | 106 |
| $50,000–$69,999 | 15 | 13 | 17 | 4 | 0 | 49 |
| $70,000 and over | 13 | 15 | 33 | 7 | 2 | 70 |
| **Totals** | 67 | 100 | 127 | 45 | 12 | 351 |

**Source: data analysis**

The distribution of trust of strangers on the street with examined with respect to the geographic region of the United States where respondents lived, the data of which are presented in Table 10. In general, distrust of strangers appears to be concentrated mainly on the east coast of the United States, and mostly in the Middle Atlantic and South Atlantic states (20.5% of grand total, 43.1% of respondents that definitely and somewhat distrust strangers). While the least distrust of strangers on the street was observed for the New England states, this region was also the least trusting of strangers. The greatest concentration of indifference toward strangers appears to be concentrated in the South Atlantic and East South Central states. Across all geographic regions of the United States, respondents mainly distrust rather than trust strangers on the street. Chi-square analysis was performed on the data. There is no statistically significant relationship between the trust of strangers on the street and the geographic region of the United States where respondents live, $X^2(24, N = 351) = 32.11, p = .124$.

**Table 10**

*Respondent trust of strangers on the street according to US geographic region*

| Region | 1 Definitely distrust | 2 Somewhat distrust | 3 Neither trust nor distrust | 4 Somewhat trust | 5 Definitely trust | Totals |
|---|---|---|---|---|---|---|
| Middle Atlantic (NY/NJ/PA) | 10 | 18 | 20 | 17 | 2 | 67 |
| New England (CT/ME/MA/NH/RI//VT) | 6 | 4 | 5 | 1 | 0 | 16 |
| South Atlantic (DE/DC/FL/GA/MD/NC/SC/VA/WV) | 21 | 23 | 32 | 12 | 2 | 90 |
| East South Central (AL/KY/MS/TN) | 6 | 20 | 26 | 2 | 3 | 57 |
| West South Central (AR/LA/OK/TX) | 9 | 13 | 19 | 5 | 4 | 50 |
| Mountain (AZ/CO/ID/MT/NV/NM/UT/WY) | 5 | 7 | 10 | 5 | 1 | 28 |
| Pacific (AK/CA/HI/OR/WA) | 10 | 15 | 15 | 3 | 0 | 43 |
| **Totals** | 67 | 100 | 127 | 45 | 12 | 351 |

**Source: data analysis**

The assumptions for structural equation modeling (SEM) were checked. The normality of residuals requirement was met (Shapiro-Wilk = 0.995, p=0.285), as was the independence of residuals (DW=1.884, 1.5<DW<2.5). The linearity assumption was partially met as some relationships were nonlinear. The assumption of homoscedasticity was not met since variances were not constant. Finally, there was some multicollinearity detected among variables. However, although there were some violations of the assumptions, maximum likelihood estimation using robust standard errors (MLR) may still be used to estimate the model and account for these issues (Mansournia et al., 2021). Structural equation modelling using latent variable analysis (lavaan) was employed to estimate a structural equation model for the trust of strangers on the street. Table 11 summarizes the model fit indices. The model achieved convergence after 47 iterations, and the model fit indices in Table 11 indicate reasonable error (RMSEA between 0.06 and 0.08, Hu & Bentler, 1999), and acceptable fit (CFI, TLI, GFI > 0.9, Hu & Bentler, 1999; SRMR <0.08, Byrne, 1994), suggesting that the model is suitable.

**Table 11**

*Model fit indices*

| Index | Value |
|---|---|
| Estimation method | MLR (Maximum likelihood with robust standard errors) |
| Degrees of freedom | 289 |
| Model chi-square | 1247.32 |
| Baseline model chi-square | 3303.627 |
| P-value | <.001 |
| Comparative fit index (CFI) | 0.921 |
| Tucker-Lewis index (TLI) | 0.912 |
| Root mean square error of approximation (RMSEA) | 0.068 |
| Goodness of fit index (GFI) | 0.901 |
| Standardized root mean squared residual (SRMR) | 0.052 |

**Source: data analysis**

The structural equation model for the trust of strangers is illustrated in Table 12. The resulting model consists of 5 structural factors, and 26 observed variables, including the outcome variable, trust of strangers on the street. In general, the model shows that all variables in the measurement models are statistically significant (p < .000), and with the exception of information trust (p=0.307), four of the five structural equation model factors are statistically significant. The model factors were evaluated using Cronbach's alpha. Social trust (α = 0.859), person trust (α = 0.841), institutional trust (α = 0.702), information trust (α = 0.812) were all shown to be reliable and of great internal consistency (Taber, 2018). However, the factor demographics (α = 0.204) was not shown to be reliable (Taber, 2018), given the intuition that certain personality traits were expected to be good predictors of trust of strangers on the street.

Social trust has the largest positive effect on the trust of strangers on the street, followed by institutional trust. Demographic characteristics have a negligible positive effect on the trust of strangers, which is statistically significant. Person trust, which embodies the trust of family and friends, has a mild negative effect on the trust of strangers, and is statistically significant. Curiously, the negative coefficient for person trust implies that as the trust of family and friends increases, the trust in strangers decreases. Finally, information trust is not statistically significant, and has a negligible positive effect on the trust of strangers.

**Table 12**

Structural equation model coefficients

| Path | Estimate | Significance | Std. Error | z-value | P-value |
|---|---|---|---|---|---|
| Person Trust | -0.106 | * | 0.045 | -2.356 | 0.018 |
| Demographics | 0.099 | * | 0.042 | 2.357 | 0.018 |
| Institutional Trust | 0.254 | * | 0.048 | 5.292 | 0 |
| Social Trust | 0.602 | *** | 0.051 | 11.804 | 0 |
| Information Trust | 0.047 |  | 0.046 | 1.022 | 0.307 |

Note: *p < .05, **p < .01, ***p < .001

Table 13 illustrates the coefficients and significance levels for the variables that feed each measurement model. All constructs and variables are statistically significant. Among institutional trust variables, the top three loadings were trust in business owners, teachers, and doctors, respectively. Among social trust variables, the top three loadings were trust in commercial actors, social media influencers, and celebrities, respectively. Among person trust variables, trust in family was slightly greater than trust in friends. Among information trust variables, the top three loadings were trust of Google business listings, trust of information from an internet search, and business websites, respectively. Finally, among demographic variables, while level of income, education, and age appear to positively influence trust in strangers, the variable everlied, representing if a respondent has ever posted a false online review, has a negative influence on trust in strangers, although with weak significance. Gender and region of residence in the United States also appear to have weak statistical significances.

**Table 13**

Measurement model coefficients

| Construct | Variable | Loading | Significance | Std.Error | z-value | P-value |
|---|---|---|---|---|---|---|
| Person Trust | Family | 0.89 | *** | 0.024 | 37.083 | 0 |
| Person Trust | Friends | 0.87 | *** | 0.024 | 36.25 | 0 |
| Institutional Trust | Doctors | 0.71 | *** | 0.028 | 25.357 | 0 |
| Institutional Trust | Politicians | 0.68 | *** | 0.029 | 23.448 | 0 |
| Institutional Trust | Religious | 0.65 | *** | 0.03 | 21.667 | 0 |
| Institutional Trust | Businessowners | 0.77 | *** | 0.026 | 29.615 | 0 |
| Institutional Trust | Teachers | 0.73 | *** | 0.027 | 27.037 | 0 |
| Social Trust | Salespeople | 0.78 | *** | 0.026 | 30 | 0 |
| Social Trust | Celebrities | 0.82 | *** | 0.024 | 34.167 | 0 |
| Social Trust | Commercialactors | 0.85 | *** | 0.023 | 36.957 | 0 |
| Social Trust | Socmediainfluencers | 0.83 | *** | 0.024 | 34.583 | 0 |
| Social Trust | Newimmigrants | 0.71 | *** | 0.028 | 25.357 | 0 |
| Information Trust | Buswebsite | 0.81 | *** | 0.024 | 33.75 | 0 |
| Information Trust | BusFacebookpage | 0.79 | *** | 0.025 | 31.6 | 0 |
| Information Trust | Periodicals | 0.75 | *** | 0.027 | 27.778 | 0 |
| Information Trust | Internetsearch | 0.82 | *** | 0.024 | 34.167 | 0 |
| Information Trust | Googlebuslisting | 0.84 | *** | 0.023 | 36.522 | 0 |
| Information Trust | Internetdiscussforum | 0.77 | *** | 0.026 | 29.615 | 0 |
| Information Trust | consumerratingssite | 0.8 | *** | 0.025 | 32 | 0 |
| Demographics | Age | 0.45 | ** | 0.037 | 12.162 | 0.002 |
| Demographics | Gender | 0.38 | * | 0.039 | 9.744 | 0.013 |
| Demographics | Income | 0.52 | ** | 0.035 | 14.857 | 0.003 |
| Demographics | Education | 0.49 | ** | 0.036 | 13.611 | 0.002 |
| Demographics | Region | 0.35 | * | 0.039 | 8.974 | 0.015 |
| Demographics | Everlied | -0.41 | ** | 0.038 | -10.789 | 0.004 |

Note: *p < .05, **p < .01, ***p < .001

**Source: data analysis**

Figure 2 illustrates the resulting structural model with the respective coefficients for each variable and factor.

## Figure 2

*Structural equation model path diagram*

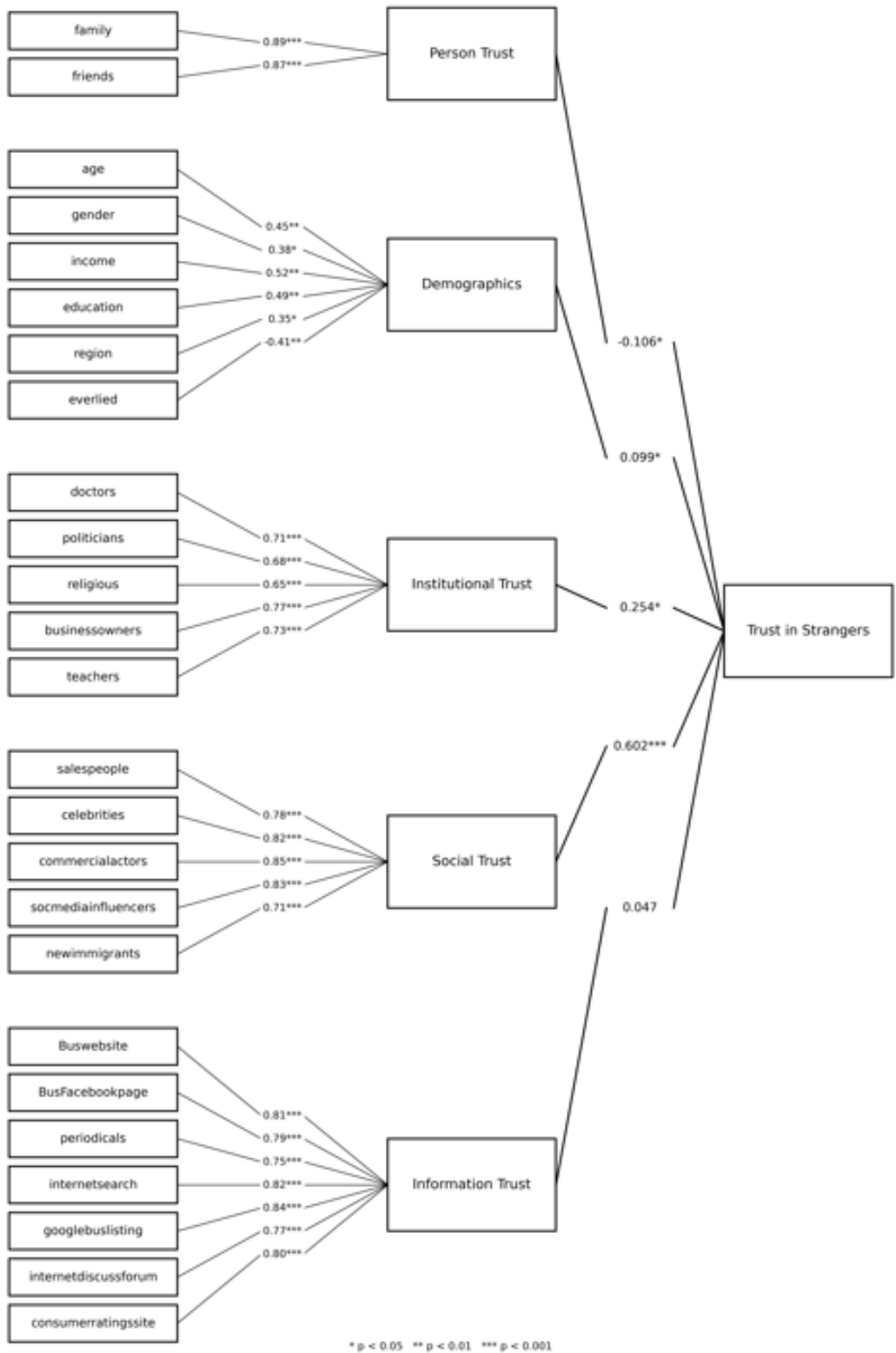

* p < 0.05   ** p < 0.01   *** p < 0.001

Note: *p < .05, **p < .01, ***p < .001

**Source:** data analysis

## 4. Discussion

In general, while 16.24% of respondents somewhat and definitely trust strangers on the street, 47.58% of respondents somewhat and definitely distrust strangers on the street. The level of distrust observed in this study echoes the observed by the Pew Research Center the same percentage of people that declared "most people can't be trusted" (Pew Research Center, 2019). However, given that 36.18% of respondents neither trust nor distrust strangers on the street, this implies that strangers are not explicitly trusted. Based on this observation, it can be concluded, therefore, that the majority of respondents do not trust strangers. The chi-square analysis revealed that the trust of strangers on the street has statistically significant relationships to certain demographic characteristics, namely age, income level, and level of education, confirming observations by Almakaeva, Welzel, & Ponarin (2018), Mirfardi (2011), Ermisch et al. (2009), and Alesina and La Ferrara (2002). The observations of this stud also generally confirm those of Edelman (2019) with regard to income, education and age characteristics of low and high trusting individuals. The observation by Ermisch et al. (2009) that older people trust others more was not supported by the data in this study. However, the chi-square analysis did not find statistically significant relationships for the trust of strangers on the street versus gender or region of residence in the United States. Interestingly, the variable everlied, a proxy for honesty was also statistically significant in chi-square analysis, and had a statistically significant negative effect on the trust of strangers in structural equation modeling. This observation suggests that the propensity to post untruthful information in online reviews may indicate a disregard for unknown others that will read the posts, and therefore, a lack of trust of strangers.

Structural equation modeling using the conceptual framework of the trust of strangers on the street revealed that while social trust, person trust, and institutional trust were all shown to be reliable with respect to internal consistency (α between 0.702 and, 0.859), only information trust was not shown to be as reliable in this model despite having a Cronbach of 0.92 in Berry (2024) as an instrument. However, the role of information trust as a source of influence for shaping opinions or perceptions of strangers cannot be discounted, given that people rely heavily on opinions and advice from random strangers online. Structural equation modeling revealed that social trust and institutional trust both had the largest positive effect on the trust of strangers, respectively, confirming the observation of Kaina (2011). The role of close relationships with friends and family, as expressed by the person trust construct, suggests a negative effect on the trust of strangers on the street. This observation implies that people are perhaps less willing to trust unknown others if there is already a strong gravitational pull from those in their personal inner circle. Finally, while most of the variables were statistically significant in the chi-square analysis, the role of demographics as a factor was also shown to be statistically significant in the structural model but with a negligible positive effect on the trust of strangers.

The loadings of the variables on each factor should serve to inform which forms of influence are the strongest to activate positive trust of strangers on the street. The largest loadings in Table 13 appear to be from the social trust and information trust factors. The largest loading from Table 13 was Google business listing from the information trust factor, illustrating the importance of online information sources. The role of people as influencers is evidenced by the large loadings for social media influencers and celebrities. Among institutional trust loadings, business owners, teachers and doctors had the highest loadings. These observations indicate that these people are agents with the greatest effect for social change with respect to developing positive trust toward strangers, whether through charity or directly helping unknown others.

These observations have implications for business and also those that want to enable social change in general. First, since it is clear that many people are generally distrustful of strangers, particularly young people and those with lower income and education levels, the root cause of this distrust must be addressed. The problem with general distrust is that it may suggest that those with low trust are less likely to lend assistance in an emergency situation and support causes that help society perhaps due to their own personal circumstances or maybe pessimism. Therefore, businesses as agents of social change that are trusted by people should take a greater role in social change by inspiring these people to want to help others and create a kinder society, and therefore develop greater trust in unknown others. From a marketing standpoint, businesses can earn more loyalty if their approach is authentic and the desire to help society is encoded into their mission and corporate culture to be effective. From a social standpoint, since businesses have greater trust than governments as observed by Edelman (2021), we should expect those organizations to want to step up and be models for social change, whether through inspiring consumers, empowering employees to be change agents, or creating some impetus that causes more kindness in society. Second, given that over a third of respondents reported to neither trust nor distrust strangers on the street, this observation is cause for concern. That respondents neither trust nor distrust strangers suggests either a certain apathy or perhaps a passive or latent dislike for strangers that the respondents choose not to explicitly declare, implying that perhaps they actually distrust strangers. Marketers and those in charities or other NGOs should look closely at why people may hold such sentiments and determine how to overcome this apparent apathy or negative perception of strangers. The problem with this apathy is that it suggests an unwillingness to help, and the indifference toward others implies that there also might be an unwillingness to help charitable causes or to assist with solving social problems, like hunger or homelessness, for example. Therefore, finding the levers to undo apathy and foster a caring society is required. The agents of change identified in this study can best reach those people that neither trust nor distrust strangers with the appropriate messaging that is relatable.

## 5. Directions for future research

Future research should focus on cues that might reveal true trust feelings toward strangers among those people that express ambiguous "neither trust nor distrust" attitudes. Given that a substantial number of respondents expressed this sentiment, the development of supplemental instruments may be required to test if respondents are harboring latent antisocial tendencies or their actual attitudes toward others or are simply being untruthful. Questions that require yes-no answers about attitudes or perceptions may be helpful to confirm trust if a respondent is forced to pick an answer (example: do you like homeless people?) when given a level of trust (example: how much do you trust homeless people?). Identifying the levers that activate a change from apathy toward strangers to caring or wanting to help should be a priority. The result of this inquiry should be to increase giving to charities and the desire to help and care for unknown others.

## 6. Limitations

In this study, as with all studies, thoughtful consideration of possible shortcomings and limitations must be given. First, as acknowledged in Berry (2024), despite its convenience, Amazon Mechanical Turk is not without shortcomings, such as the potential for gamification by users to qualify for surveys. Since measures were taken to ensure data quality, such as attention check questions, the dataset was cleaned to exclude users that fail these checks. Therefore, researchers should use mTurk with caution as the sample will likely be reduced as a result. Second, the constructs used in this study were based on the trust attitudinal data that was collected in Berry (2024). The selection of items used for the constructs were related to everyday sources of personal influence in the context of purchase decision making in my dissertation, whether by people, authority figures or media sources, and therefore, were not meant to be exhaustive.

## 7. Patents

There are no patents resulting from the work reported in this manuscript.

## 8. Funding

This research received no external funding.

## 9. Conflicts of Interest

The authors declare no conflict of interest.

## 10. Declaration of generative AI in scientific writing

The author declares that generative AI tools were not used in the writing or research of this article.